\documentclass[10pt]{llncs} 
\pdfoutput=1                    

\usepackage{xspace, enumerate, boxedminipage}
\usepackage{algorithm,algorithmic}
\usepackage{mathtools,amsmath, amssymb, amsfonts }
\usepackage{graphicx}
\usepackage{sober,lcsect}

\usepackage[usenames,dvipsnames]{color}

\newcommand{\parasec}[1]{\vspace{8px}\noindent\textbf{#1}\hspace{5px}}
\newcommand{\ie}{\emph{i.e.\,}}
\newcommand{\etal}{\emph{et al.\,}}
\newcommand{\F}{\mathbb{F}}
\newcommand{\whp}{w.h.p.\xspace}
\DeclareMathOperator{\polylog}{polylog}
\DeclareMathOperator{\poly}{poly}
\newcommand{\fbad}{\frac18\xspace}
\newcommand{\fbadq}{$q/8$\xspace}
\newcommand{\fgoodq}{7/8\xspace}
\newcommand{\fgoodqq}{$\frac{7q}{8}$\xspace}
\newcommand{\thresh}{5/8\xspace}
\newcommand{\threshq}{$\frac{5q}{8}$\xspace}
\newcommand{\dcl}{D}
\newcommand{\flag}{$\langle\mathsf{Flag}\rangle$\xspace}

\newcommand{\countm}{$\langle\mathsf{Count}\rangle$\xspace}
\newcommand{\done}{$\langle\mathsf{Done}\rangle$\xspace}

\newcommand{\oneinputs}{$\mathsf{1}$-inputs\xspace}

\newcommand{\vssSh}{\textsc{Avss-Share}\xspace}
\newcommand{\vssRec}{\textsc{Avss-Rec}\xspace}
\newcommand{\hw}{\textsc{Hw-MPC}\xspace}
\newcommand{\ba}{\textsc{Asynch-BA}\xspace}

\newcommand{\tc}{\textsc{$\tau$-Counter}\xspace}
\newcommand{\cq}{\textsc{Create-Quorum}\xspace}
\newcommand{\ic}{\textsc{Input-Commitment}\xspace}
\newcommand{\icPlayer}{\textsc{IC-Player}\xspace}
\newcommand{\icInput}{\textsc{IC-Input}\xspace}
\newcommand{\ce}{\textsc{Circuit-Eval}\xspace}
\newcommand{\gm}{\textsc{Mask-Generation}\xspace}
\newcommand{\gcomp}{\textsc{Gate-Eval}\xspace}
\newcommand{\outrec}{\textsc{Output-Rec}\xspace}
\newcommand{\outprop}{\textsc{Output-Propagation}\xspace}
\newcommand{\maj}{\thresh-\textsc{Majority}\xspace}
\newcommand{\outnode}{output node}
\newcommand{\outval}{\langle\mathsf{O}\rangle\xspace}
\newcommand{\ct}{count tree\xspace}

\DeclareMathOperator{\vl}{\mathsf{y}}
\DeclareMathOperator{\mask}{\mathsf{r}}
\DeclareMathOperator{\mv}{\mathsf{\hat{y}}}
\DeclareMathOperator{\lef}{\mathsf{left}}
\DeclareMathOperator{\rit}{\mathsf{right}}
\DeclareMathOperator{\outnodemath}{\text{output-node}}

\newcommand{\varsha}[1]{\textcolor{blue}{$\bullet$}\marginpar{\textcolor{blue}{{\bf Varsha:} #1}}}

\newcommand{\tourn}{election}
\newcommand{\enode}{e-node}
\newcommand{\ane}{an}
\newcommand{\toplev}{\ell^*}

\newcommand{\lev}{\ell}

\newcommand{\subcomm}{{\sc Elect-Subcommittee}}

\newcommand{\heavyba}{{byzantine agreement}\xspace}
\newcommand{\lst}{\ensuremath{\mathsf{List}}}
\newenvironment{indentpar}[1]%
 {\begin{list}{}%
         {\setlength{\leftmargin}{#1}}%
         \item[]%
 }
 {\end{list}}
 
 \newcommand{\rs}{semi-random-string agreement\xspace}
 \newcommand{\binelection}{{bin election}\xspace}
 \newcommand{\es}{ELECT-SUBCOMMITTEE\xspace}
 \newcommand{\rsAlg}{SRS-AGREEMENT\xspace}
 \newcommand{\rsToQ}{SRS-TO-QUORUM\xspace}
\begin{document}
\title{Quorums Quicken Queries: Efficient Asynchronous Secure Multiparty Computation}
\author{
Varsha Dani\inst{1}
\and Valerie King\inst{2} 
\and Mahnush Movahedi\inst{1}
\and Jared Saia\inst{1}\thanks{Partially supported by NSF
CAREER Award 0644058 and NSF CCR-0313160.}
}
\institute{University of New Mexico 
\and University of Victoria}
\maketitle



\begin{abstract}
We describe an asynchronous algorithm to solve secure multiparty
computation (MPC) over $n$ players, when strictly less than a $\fbad$
fraction of the players are controlled by a static adversary.  For any
function $f$ over a field that can be computed by a circuit with $m$ gates, our
algorithm requires each player to send a number of field elements and perform an
amount of computation that is $\tilde{O}(\frac{m}{n} + \sqrt n)$. This
significantly improves over traditional algorithms, which require each
player to both send a number of messages and perform computation that
is $\Omega(nm)$.

Additionaly, we define the \emph{threshold counting problem} and
present a distributed algorithm to solve it in the asynchronous
communication model.  Our algorithm is load balanced, with
computation, communication and latency complexity of $O(\log{n})$, and
may be of independent interest to other applications with a load
balancing goal in mind.
\end{abstract}

\section{Introduction}

Recent years have seen a renaissance in secure multiparty computation (MPC), but unfortunately, the distributed computing community is in danger of missing out.  In particular,  while new MPC algorithms boast dramatic improvements in latency and communication costs, none of these algorithms offer significant improvements in the highly \emph{distributed} case, where the number of players is large.  

This is unfortunate, since MPC holds the promise of addressing many important problems in distributed computing. How can peers in Bittorrent auction off resources without hiring an auctioneer?  How can we design a decentralized Twitter that enables provably anonymous broadcast of messages.  How can we create deep learning algorithms over data spread among large clusters of machines?

In this paper, we take a first step towards solving MPC for large distributed systems.  We describe algorithms that require each player to send a number of messages and perform an
amount of computation that is $\tilde{O}(\frac{m}{n} + \sqrt n)$, where $n$ is the number of players and $m$ is the number of gates in the circuit to be computed. This significantly improves over current algorithms, which require each player to both send a number of messages and perform computation that
is $\Omega(nm)$.  We now describe our model and problem.
 
\parasec{Model} There are $n$ players, with a private and authenticated channel between every pair of players.  Communication is via asynchronous message passing, so that sent messages may be arbitrarily and adversarially delayed. Latency in this model is defined as the maximum length of any chain of messages (see \cite{CD,AW}). 

We assume a Byzantine adversary controls an unknown subset of up to $t$ of the players.  These players are \emph{bad} (\ie Byzantine) and the remaining
players are \emph{good}. The good players run our algorithm, but the bad players may deviate in an arbitrary manner.  Our adversary is \emph{static}, \ie it must select the set of bad players at the start of our algorithm.  The adversary is computationally unbounded. Thus, we make no cryptographic hardness assumptions.

\smallskip
\noindent
{\bf MPC Problem:} Each player, $p_{i}$, has a private input $x_{i}$.  All players know a $n$-ary function $f$.  We want to ensure that: 1) all players learn the value of $f$ on the inputs; and 2) the inputs remain as private as possible: each player $p_{i}$ learns nothing about the private inputs other than what is revealed by the output of $f$ and the player's private value $x_i$.

In the asynchronous setting, the problem is challenging even with a
trusted third party.   In particular, the trusted party can not determine the difference between a message never being sent and a message being arbitrarily delayed, and so the $t$ bad players can always refrain from sending any messages to the trusted party.   Thus, the trusted party must wait to receive $n-t$ inputs.  Then it must compute the function $f$ using default values for the missing inputs, and send the output  back to the players as well as the number of received inputs.\footnote{We send back only the number of inputs used, not the set of players whose inputs are used. This is required to ensure scalability; see Section \ref{s:dis}.}  The goal of an asynchronous MPC protocol is to
simulate the above scenario, without the trusted third party. 

The function to be computed is presented as a circuit $C$ with $m$
gates. For convenience of presentation, we assume each gate has fan-in 
two and fan-out at most two. For any two gates $x$ and $y$ in $C$, if the 
output of $x$ 
is input to $y$, we say that $x$ is a \emph{child} of $y$ and
that $y$ is a \emph{parent} of $x$. We also assume that all
computations in the circuit occur over a finite field $\F$; The size of 
$\F$ depends on the specific function to be computed but must always be 
$\Omega(\log{n})$. All the inputs,
outputs and messages sent during the protocol are elements of $\F$, and 
consequently, messages will be of size $\log|\F|$.


\smallskip
\noindent
Our MPC result requires solutions to the following two problems, which may be of independent interest.

\smallskip
\noindent
{\bf Threshold Counting}  There are $n$ good players each 
with a bit initially set to $\mathsf{0}$. At least $\tau$ of the
players will eventually set their bits to $\mathsf{1}$. The goal is for 
all the players to learn when the number of bits with values $\mathsf{1}$ is at least $\tau$. 

\smallskip
\noindent
{\bf Quorum Formation} There are $n$ players, up to $t$ of whom 
may be bad.
A \emph{quorum} is a set of $c \log{n}$ players for some constant $c$.  
A quorum is called \emph{good} if the fraction of bad players in it is 
at most $t/n+\delta$ for a fixed positive $\delta$. We want all $n$ players to agree on a set of $n$ good quorums, and we want the quorums to be load-balanced: each
player is mapped to $O(\log n)$ quorums.



\subsection{Our Results}


The main result of this paper is summarized by the following theorem.

\begin{theorem} \label{theo:1} 
Assume there are $n$ players, less than a $\fbad-\epsilon$ fraction of which
are bad for some fixed $\epsilon>0$, and an $n$-ary function, $f$ that can be computed by a circuit 
of depth $d$ with $m$ gates. If all good players follow 
Algorithm~\ref{alg:main}, then with high probability (\whp), they will solve MPC, while ensuring:
\begin{enumerate}
\item Each player sends at most $\tilde{O}(\frac{m}{n} + \sqrt n)$
  field elements,
\item Each player performs $\tilde{O}(\frac{m}{n}+ \sqrt n)$ computations, and
\item Expected total latency is $O(d \polylog(n))$.
\end{enumerate}
\end{theorem}

\noindent
Our additional results are given by the following two theorems.

\begin{theorem}
Assume $n$ good players follow Algorithm \tc.   Then \whp, 
the algorithm solves the threshold counting problem, while ensuring:
\begin{enumerate}
\item Each player sends at most $O(\log{n})$ messages of constant size,
\item Each player receives at most  $O(\log{n})$ messages,
\item Each player performs $O(\log{n})$ computations,
\item Total latency is $O(\log{n})$.
\end{enumerate}
\label{theo:2}
\end{theorem}

\begin{theorem}
Assume $n$ players, up to $t<(\frac14-\epsilon)n$ of whom are 
bad, for fixed $\epsilon>0$.  If all good players follow the \cq protocol, the following are ensured \whp:
\begin{enumerate}
\item The players agree on $n$ good quorums,
\item Each player sends at most $\tilde{O}(\sqrt n)$ bits,
\item Each player performs $\tilde{O}(\sqrt n)$ computations,
\item Total latency is $O(\polylog(n))$.
\end{enumerate}
\label{thm:quorum-formation}
\end{theorem}

In the rest of the paper we discuss the algorithms and ideas involved in 
obtaining these results. Detailed proofs are deferred to the full 
version~\cite{dkms2013}.

\section{Related Work}

The study of secure computation started in 1982 with the seminal work
of Yao~\cite{yao1982protocols}. Later Goldrich, Micali, and
Wigderson~\cite{goldreich1987play} proposed the first generic scheme
for solving a cryptographic notion of MPC. This work was followed by
some unconditionally-secure schemes in late 1980s~\cite{benor_goldwasser_wigderson:completeness,chaum_crepeau_damgard:multiparty,rabin1989verifiable,benor_canetti_goldreich:asynchronous,hirt2001robustness,hirt2005upper,ZH06}. 
Unfortunately, these methods all have poor communication scalability that prevents their 
wide-spread use. In particular, if there are $n$ players
involved in the computation and the function $f$ is represented by a
circuit with $m$ gates, then these algorithms require each player to
send a number of messages and perform a number of computations that is
$\Omega(mn)$ (see~\cite{frikken2010secure,goldreich1998secure,du2001secure}).

Recent years have seen exciting improvements in the cost of MPC when
$m$ is much larger than $n$
\cite{damgard2006scalable,damgard2007scalable,damgard2008scalable}. 
For example, the computation and communication cost for the algorithm described by Damg{\aa}rd \etal
in
~\cite{damgard2008scalable} is $\tilde{O}(m)$ plus a polynomial in
$n$. However, the additive polynomial in $n$ is large
(e.g. $\Omega(n^{6})$) and so these new algorithms are only efficient
for relatively small $n$. Thus, there is still a need for MPC
algorithms that are efficient in both $n$ and $m$.

We first introduced the notion of using quorums for local
communication to decrease the message cost in a brief
announcement~\cite{Dani12}. In that paper, we described a synchronous
protocol with bit complexity of $\tilde{O}(\frac{m}{n} + \sqrt{n})$
per player that can tolerate a computationally unbounded adversary who
controls up to $(\frac14 - \epsilon)$ fraction of the players for any
fixed positive $\epsilon$. This paper improves our previous result by
handling asynchronous communication. One important challenge in 
the asynchronous communication model is to ensure that at least 
$n-t$ inputs are committed to, before the circuit evaluation. 
To address this issue we introduce and solve the \emph{threshold counting 
problem.}

Boyle, Goldwasser, and Tessaro~\cite{Boyle13} describe a synchronous
cryptographic protocol to solve MPC problem that is also based on
quorums. Their algorithm uses a fully homomorphic encryption (FHE)
scheme and thus, tolerates a computationally-bounded adversary that
can take control of up to ($\frac13 - \epsilon$) fraction of players for
any fixed positive $\epsilon$. Their protocol requires each player to
send $\polylog(n)$ messages of size $\tilde{O}(n)$ bits and
requires $\polylog(n)$ rounds. Interestingly
the cost of the protocol is independent of the circuit size.

\smallskip
\noindent
{\bf Counting Networks} 
Threshold counting can be solved in a 
load-balanced way using
\emph{counting networks}, which were first introduced
by Aspnes, Herlihy, and Shavit~\cite{Aspnes:1991:CNM:103418.103421}.
Counting networks are
constructed from simple two-input two-output computing elements called
\emph{balancers} connected to one another by wires. A counting network
can count any number of inputs even if they arrive at arbitrary times,
are distributed unevenly among the input wires, and propagate through
the network asynchronously. Aspnes, Herlihy, and 
Shavit~\cite{Aspnes:1991:CNM:103418.103421} establish an $O(\log^2 n) $
upper bound on the depth complexity of counting networks.  Since the
latency of counting is dependent to the depth of the network,
minimizing the network's depth is a goal for papers in this area. A
simple explicit construction of an $O(\log{n} c^{ \log^*{n}})$-depth
counting network, and a randomized construction of an $O(\log n)
$-depth counting network
which works with high probability is described in
\cite{Klugerman:1992:SCN:129712.129752,Klugerman94small-depthcounting}. 
These constructions use the AKS sorting
network~\cite{Ajtai:1983:SN:800061.808726} as a building block.
While the AKS sorting network and the resulting counting networks
have $O(\log{n})$ depth, large hidden constants render them impractical.
We note that the threshold counting problem is simpler than general 
counting.

\section{Preliminaries}\label{s:pre}

We say an event occurs \emph{with high probability
  (w.h.p)}, if it occurs with probability at least $1-1/n^c$, for some
$c>0$ and sufficiently large $n$. We assume all computations occur
over a finite field $\F$. Every time we use a mask during the
protocol, we assume the mask is a value chosen uniformly at random
from $\F$.

We now describe protocols that we use as building
blocks in this paper.

\smallskip
\noindent
{\bf Secret Sharing} In secret sharing, a player, called the
dealer, wants to distribute a secret amongst a group of participants,
each of whom is allocated a share of the secret. The secret can be
reconstructed only when a sufficient number of shares are combined
together and each of the shares reveals nothing to the player
possessing it. If a method is used to ensure the dealer sends shares
of a real secret and not just some random numbers, then the new scheme
is called Verifiable Secret Sharing (VSS). As our model is
asynchronous, we use the asynchronous VSS (or AVSS) scheme described
by Benor, Canneti and Goldreich
in~\cite{benor_canetti_goldreich:asynchronous}. We denote the sharing
phase by \vssSh and the reconstruction phase by \vssRec. The protocol
of ~\cite{benor_canetti_goldreich:asynchronous} works correctly even
if up to $\frac14$ of the players are bad. The latency of the
protocols is $O(1)$ and the communication cost is $\poly(q)$, where
$q$ is the number of players participating in the protocol. In this
paper, we will use the protocols only among small sets of players
(quorums) of logarithmic size, so $q$ will be $O(\log{n})$ and the
communication cost per invocation will be $\polylog(n)$.

\smallskip
\noindent
{\bf Heavy-Weight MPC} We use a heavy-weight asynchronous
algorithm for MPC donated by \hw. This algorithm, due to Ben-Or et
al.~\cite{benor_canetti_goldreich:asynchronous}, is an errorless MPC
protocol that tolerates up to $\frac14$ bad players. Let $q$ be the number
of players who run a \hw to compute a circuit with $O(q)$ gates. The
expected latency of \hw is $O(q)$ and the number of messages 
sent $\poly(q)$. In this paper, we will use \hw only for
logarithmic number of players and gates, \ie, $q = O(\log{n})$ and the
communication cost per invocation is $\polylog(n)$.\footnote{To make sure 
our algorithm has the expected total latency equal to $O(d \polylog(n))$, 
every time we need to run the \hw algorithm, we run $O(\log{n})$ same 
copy of it each for $O(\polylog(n))$ steps.}

\smallskip
\noindent
{\bf Asynchronous Byzantine Agreement} In the Byzantine agreement
problem, each player is initially given an input bit. All good players
want to agree on a bit which coincides with at least one of their
input bits. Every time a broadcast is required in our protocol, we use
an asynchronous Byzantine agreement algorithm from~\cite{canetti}, which we
call \ba.

\section{Technical Overview}

We briefly sketch the ideas behind our three results.

\begin{figure}[t]
 \begin{center}
 \includegraphics[scale=0.30]{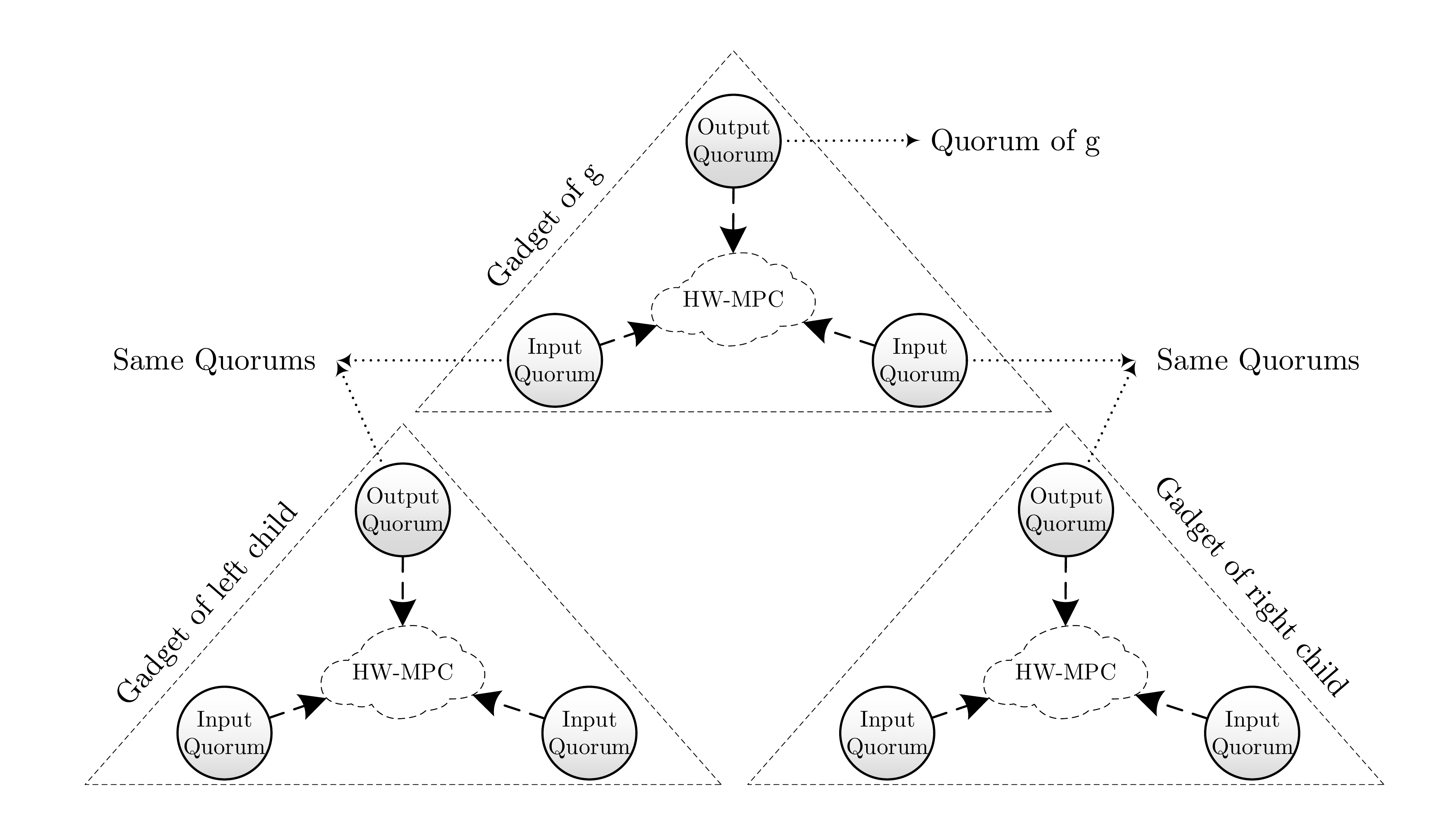}
 \end{center}
 \caption{The gate gadgets for gate $g$ and its left and right children.}
 \label{f:gadget}
 \end{figure}

\smallskip
\noindent
{\bf Quorum-Based Gate Evaluation} The main idea for reducing the
amount of communication required in evaluating the circuit is
quorum-based gate evaluation. Unfortunately, if each player
participates in the computation of the whole circuit, it must
communicate with all other players. Instead, in quorum-based gate
evaluation, each gate of the circuit is computed by a \emph{gate
  gadget}. A gate gadget consists of three quorums: two \emph{input
  quorums} and one \emph{output quorum}. Input quorums are associated
with the gate's children which serve inputs to the gate. Output quorum
is associated with the gate itself and is responsible to create a
shared random mask and maintain the output of the quorum for later use
in the circuit. As depicted in Figure~\ref{f:gadget}, these gate
gadgets connect to form the entire circuit. In particular, for any
gate $g$, the output quorum of $g$'s gadget is the input quorum of the
gate gadget for all of $g$'s parents (if any).

The players in each gate gadget run \hw among themselves to perform
the gate operation. To make sure the computation is correct and
secure, each gate gadget maintains the invariant that the value
computed by the gadget is the value that the corresponding gate in the
original circuit would compute, masked by a uniformly random element
of the field. This random number is not known to any individual
player. Instead, shares of it are held by the members of the output
quorum.  Thus, the output quorum can participate as an input quorum
for the evaluation of any parent gate and provide the masked version
of the inputs and shares of the mask.

This gate gadget computation is continued in the same way for all
gates of the circuit until the final output of whole circuit is
evaluated.  This technique for evaluating a gate of the circuit using
quorums, is illustrated in Figure~\ref{sec:compute} and the details
are described in Section~\ref{sec:count}.  

\smallskip
\noindent
{\bf Threshold Counting}
Our interest in threshold counting for this paper is to ensure that at
least $n-t$ inputs are committed to, before the circuit evaluation occurs. 
To solve the threshold counting problem, we design a new distributed data 
structure and algorithm
called \tc. The \tc enables threshold counting with 
asynchronous communication, and may be of use for other problems beyond MPC. 

To give intuition, we first consider a naive approach
for counting in asynchronous model. Assume a complete binary tree where 
each player sends its input to a unique leaf node when it is set to $1$. 
Then, for every node $v$, each child of $v$ sends $v$ a message showing 
the number of inputs it has received so far and it sends the updated 
message every time this number changes. The
problem with this approach is that it is not load-balanced: each node
at depth $i$ has $\frac{n}{2^i}$ descendants in the tree, and
therefore, in the worst case, sends and receives $\frac{n}{2^i}$
messages. Thus, a child of the root sends $n/2$ messages to the root
and receives the same number of messages from its children. To solve
the load-balancing problem, we use a randomized approach which ensures 
\whp that each leaf of the data structure receives at least $7\log n$ 
messages and does not communicate with its parent until it has done so.
Subsequnt messages it receives are not forwarded to its parent but rather 
to other randomly chosen leaves to ensure a close to uniform distribution 
of the messages. The details of our algorithm are described in 
Section~\ref{sec:count} and more formally in
Algorithm~\ref{alg:tc}. Theorem~\ref{theo:2} describes the resource complexity 
of the algorithm.


\smallskip
\noindent
{\bf Asynchronous Quorum Formation} 
 Recently, King et al.~\cite{ICDCN11} described an efficient algorithm
 to solve the quorum formation problem, \whp, in the synchronous model
 with full information. Our new algorithm, \cq, builds on the result
 of~\cite{ICDCN11} to solve the quorum formation problem in the
 asynchronous model, with private channels. The properties of \cq, are
 described by Theorem~\ref{thm:quorum-formation}. The algorithm and
 the proof  are deferred to the full version~\cite{dkms2013} due to space 
restrictions.

\section{Our Algorithm} \label{sec:alg}
Our algorithm makes use of a \emph{circuit graph}, $G$, which is based
on the circuit $C$ that computes $f$. We assume the gates of the
circuit $C$ are numbered $1, 2, \dots, m$, where the gate numbered 1
is the output gate. The circuit graph is a directed acyclic graph over
$m+n$ nodes.  There are $n$ of these nodes, one per player, that we
call \emph{input nodes}.  There are $m$ remaining nodes, one per gate,
that we call \emph{gate nodes}. For every pair of gate nodes $x$ and
$y$, there is an edge from $x$ to $y$ iff the output of the gate
represented by node $x$ is an input to the gate represented by node
$y$.  Also, for any input node $z$ and gate node $y$, there is an edge
from $z$ to $y$ if the player represented by gate node $z$ has an
input that feeds into the gate represented by node $y$. Similar to our
definition in $C$, for any two nodes $x$ and $y$ in $G$, if $x$ has an
edge to $y$, we say that $x$ is a \emph{child} of $y$ and that $y$ is
a \emph{parent} of $x$.  Also, for a given node $v$, we will say the
\emph{height} of $v$ is the number of edges on the longest path from
$v$ to any input node in $G$. For each node in $G$, we define the following variables. $Q_v$ is the quorum associated with node $v$. $y_v$ is the output of the gate corresponding to $v$. Finally, $r_v$ is a random mask and $\hat y_v$ is the masked output associated with node $v$, \ie ~$\hat y_v = y_v + r_v$.

We number the nodes of $G$ canonically 
in such a way that the input node numbered $i$
corresponds to player $p_i$. We refer to the node corresponding to the
output gate as the \emph{\outnode}.

Algorithm~\ref{alg:main} consists of four parts. The first part is to
run \cq in order to agree on $n$ good quorums. The second part of the
algorithm is \ic in which, quorums form the \ct. Then, each player
$p_i$ \emph{commits} its input values to quorum $i$ at the leaf nodes
of the \ct and finally the players in that quorum decide whether these
values are part of the computation or not. The details of this part of
the algorithm is described in Section~\ref{sec:count}. The third part
of the algorithm is evaluation of the circuit, described in detail in
Section~\ref{sec:compute}). Finally, the output from the circuit
evaluation is sent back to all the players by quorums arranged in a complete 
binary tree.

\begin{algorithm}
\caption{Main Algorithm}\label{alg:main}
\begin{enumerate}
\item All players run \cq,
\item All players run \ic,
\item All players run \ce,
\item Propagate the output by quorums arranged in a complete binary tree.
\end{enumerate}
\end{algorithm}

\subsection{\ic} \label{sec:count}
In this section we describe a Monte Carlo algorithm,
called \tc that performs threshold counting for a threshold $\tau \ge n/2$. 

The algorithm consists of up and down stages. For the up stage
the players are arranged in a (pre-determined)  tree data structure 
consisting of a root node with $O(\log{n})$ children, each of which is 
itself the root of a complete binary tree; these subtrees have 
varying depths.  The players in the trees count the number of \oneinputs, 
\ie the number of players' inputs that are set to $\mathbf{1}$. As a
result, the root can decide when the threshold is reached. In the down stage, 
the root notifies all the players of this event via a complete 
binary tree of depth $\log n$. Note that the trees used in the up and down 
stages have the same root. In what follows, unless otherwise specified, 
``tree'' will refer to the  tree data structure used for the up stage.

Let $\dcl = \lceil \log{\frac{\tau}{14\log{n}}}\rceil$. Note that 
$\dcl = O(\log{n})$. The root of our tree has degree $\dcl$. Each of 
the $\dcl$ children of the root is itself the root of a complete 
binary subtree, which we will call a \emph{collection subtree}. 
For $1\le j\le D$, the $j$th collection subtree has depth $\dcl+1-j$. 
Player 1 is assigned to the root and players 2 to $D+1$, are assigned to its 
children, \ie the roots of the collection subtrees, with player $j+1$ being 
assigned to the $j$th child. The remaining nodes of the collection trees are 
assigned players in order, starting with $D+2$, left to right and top to 
bottom. One can easily see that the entire data structure has fewer than 
$n$ nodes, (in fact it has fewer than $\frac{\tau}{3\log{n}}$ nodes) so 
some players will not be assigned to any node. 

The leaves of each collection subtree are \emph{collection nodes} 
while the internal nodes of each collection tree are
\emph{adding nodes}. 

When a player's input is set to $\mathbf{1}$, it sends a 
\flag message, which we will sometimes simply refer to as a flag, 
to a uniformly random collection node from the first collection subtree. 
Intuitively, we want the flags to be distributed close to evenly among 
the collection nodes.
The parameters of the algorithm are set up so that \whp each 
collection node receives at least $7\log{n}$ \flag messages.

Each collection node in the $j$th collection tree waits until it has 
received $7\log{n}$ flags.
It then sends its parent a \countm message. For each additional flag 
received,  up to $14 \log n$, it chooses a uniformly random collection 
node in the $(j+1)$st collection subtree and forwards a flag 
to it. If $j = \dcl$ then it forwards these $14\log{n}$ flags 
directly to the root.
Subsequent flags are ignored. Again, we use the randomness to 
ensure a close to even distribution of flags \whp

Each adding node waits until it has received a \countm 
message from each of its children. Then it sends a \countm message 
to its parent. We note that
each adding node sends exactly one message during the algorithm. 
The parameters of the algorithm are arranged so that all the \countm 
messages that are sent in the the $j$th collection subtree together account 
for $\tau/2^j$ of the \oneinputs. Thus all the \countm messages in all the 
collection subtrees together account for $\tau \left(1-\frac1{2^{\dcl}}\right)$ 
of the \oneinputs. At least $\frac{\tau}{2^{\dcl}}$ \oneinputs 
remain unaccounted for. These and upto $O(\log{n})$ more are collected 
as flags at the root.

When player 1, at the root,  has accounted for at least $\tau$ 
\oneinputs, it starts the down stage by sending the \done message 
to players 2 and 3. For $j>1$, when player $j$ receives the \done message, 
it forwards to players $2j$ and $2j +1$. Thus, eventually the  \done
message reaches all the players, who then know that the threshold 
has been met. The formal algorithm is shown in Algorithm~\ref{alg:tc}.


\begin{algorithm}
\caption{\tc}
$n$ is the number of players,  $\tau$ is the threshold. 
$\dcl = \lceil \log(\frac{\tau}{14\log{n}})\rceil$
\begin{enumerate}
\item Setup (no messages sent here):
\begin{enumerate}
\item Build the data structure: 
\begin{itemize}
\item Player 1 is the root. 
\item For $1\le j \le \dcl$, player $j+1$ is a child of the root and 
the root of the $j$th collection subtree, which has depth $\dcl+1-j$. 
\item The remainder of the nodes in the collection subtrees are assigned to 
players
left to right and top to bottom, starting with player $\dcl+2$.
\end{itemize}
\item Let $sum = 0$ for the root.
\end{enumerate}
\item Up stage 
\begin{enumerate}
\item Individual Players: upon input change to $\mathbf{1}$ choose a 
uniformly random collection node $v$ from collection subtree 1 and
send a \flag to $v$.
\item Collection nodes in collection subtree $j$: 
\begin{itemize}
\item Upon receiving $7\log{n}$ {\flag}s from collection subtree $j-1$, 
if $j>1$, 
or from  individual players
if $j=1$, 
send parent a \countm message.
\item Upon subsequently receiving a \flag, send it to a uniformly random 
collection node  in collection subtree $j+1$, if $j<\dcl$.  
If $j=\dcl$ then send 
these directly to the root. Do this for up to $14 \log n$ 
flags. Then ignore all subsequent \flag messagess.
\end{itemize}     
\item Adding nodes: Upon receiving \countm  
messages from both children, send parent 
a \countm message.
\item Root:  While $sum < \tau$
\begin{itemize}
\item Upon receiving a \countm message from the root of collection subtree $j$,
increment $sum$ by $\tau/2^j$ 
\item Upon receiving a \flag message, add one to $sum$.
\end{itemize} 
\end{enumerate}
\item Down stage (now $sum \ge \tau$)
\begin{enumerate}
\item Player 1 (the root):  Send \done to Players 2 and 3, then terminate.
\item Player $j$ for $j>1$: Upon receiving \done from Player 
$\lfloor j/2 \rfloor$, forward it to Players $2j$ and $2j+1$ (if they exist) 
and then terminate.
\end{enumerate}
\end{enumerate}
\label{alg:tc}
\end{algorithm}

The algorithm \ic is based on \tc assuming that the nodes in the data 
structure are assigned quorums. We assume that the quorums have a canonical 
numbering $Q_1,Q_2,\dots Q_n$ and the role of ``player'' $i$ in \tc is 
played by 
quorum $Q_i$. When we say quorum $A$ sends a message $M$ to quorum
$B$, we mean that every (good) player in quorum $A$ sends $M$ to every
player in quorum $B$. A player in quorum $B$ is said to have received
$M$ from $A$ if it receives $M$ from at least $\fgoodq$ of the players
in $A$.
\begin{algorithm}
\caption{\ic}\label{alg:aggr}
Run the following algorithms in parallel:
\begin{enumerate}
\item Algorithm \icPlayer.
\item Algorithm \icInput.
\item Algorithm \tc with $\tau = n-t$ and with quorums as participants.
\end{enumerate}
\end{algorithm}

\begin{algorithm}
\caption{\icPlayer}\label{alg:icindividual}
{Run by player $p_i$ with input $x_i$}
\begin{enumerate}
\item $Q_{v} \leftarrow$ the quorum at the leaf node $v$ associated with input of player $p_i$ 
\item Sample a uniformly random value  from $\F$ and set $\mask_{v}$ to this value.
\item $\mv_{v} \leftarrow x_i + \mask_{v} $
\item Send $\mv_{v}$ to all the players in $Q_{v}$
\item Run \vssSh to commit the secret value $\mask_{v}$ to the players in $Q_{v}$.
\end{enumerate}
\end{algorithm}

\begin{algorithm}
\caption{\icInput}\label{alg:icinput}
{Run by player $p_j$ in Quorum $Q_v$ associated with node $v$ responsible for the input of $p_i$}
\begin{enumerate}
\item After receiving $\mv_{v}$  and a share of $\mask_{v}$,  
participate in the \vssSh verification protocol and agreement protocol 
to determine whether consistent shares for $\mask_{v}$ and the same $\mv_{v}$ are sent to everyone.
\item If the \vssSh verification protocol and the agreement protocol end and 
it is agreed that $\mv_{v}$ was the same for all and shares of $\mask_{v}$ are 
valid and consistent, set $b_{i,j} \leftarrow 1$ and you are ready to start the \tc algorithm with \flag message.
\item Upon receiving \done from your parent quorum,  participate 
in \maj  using $b_{i,j}$ as your input. If it returns FALSE, reset $\mv_{v}$ to the default value and your share of $\mask_v$ to 0.
\end{enumerate}
\end{algorithm}

The \tc is used for input commitment in the following way.
Let $v$ denote the input node associated with player $p_i$ who holds input 
$x_i$. Quorum $Q_i$ is assigned to this node. 
Player $p_i$  samples $\mask_{v}$  uniformly at random from $\F$,
sets $\mv_{v}$ to $x_{i} + \mask_{v}$ and sends
$\mv_{v}$ to all players in $Q_i$.  Next, player $p_i$ uses \vssSh to 
commit to the secret value $\mask_{v}$ to all
players in $Q_i$.  Once player $p_i$ has verifiably completed this
process for the values $x_{i}$ and $\mask_{v_i}$, we say that player
$p_i$ has \emph{committed} its masked value to $Q_i$, and each
player $p_j$ in $Q_i$ then sets a bit $b_{i,j}$ to 1. If player
$p_i$'s shares \emph{fail} the verification process, then the
$b_{i,j}$'s are set to 0, which is also the default value. 
Note that the quorum $Q_i$'s input for \tc will be $\mathbf{1}$ if \thresh 
of the 
$b_{i,j}$ are 1. The
quorums acting as nodes in the \tc data structure, run Algorithm~\ref{alg:tc} 
with threshold of $n-t$ to determine when at least
$n-t$ inputs have been committed. Note that when a quorum has to select a 
random quorum to communicate with, they must agree on the quorum via a 
multiparty computation.

Based on the down stage in \tc, when at least $n-t$ inputs have been
detected at the root node, it sends a \done message to all the quorums 
via a complete binary tree.
When a player $p_j$ who is a member of quorum $Q_i$ receives the \done 
message, $p_j$ participates
in a \hw with other members of $Q_i$, using $b_{i,j}$ as its
input. This \hw determines if at least \thresh of the bits are set to
1. If they are, then the quorum determines that the $i$-th input
($x_i$) is part of the computation and uses the received value of
$\mv_{v}$ and shares of $\mask_{v}$ as their input into
\gcomp. Otherwise, they set $\mv_{v}$ to the default input and the
shares of $\mask_{v}$ to 0. We call this the \maj step.

\subsection{Evaluating the Circuit} \label{sec:compute}

We assign nodes of $G$ to quorums in the following way. The output
node of $G$ is assigned to quorum $1$; then every node in $G$ numbered
$i$ (other than the output node) is assigned to quorum number $j$,
where $j = (i \mod n)$.  
Assume player $p$ is in quorum $Q_v$
at the leaf node $v$ of \ct, which is the same quorum assigned to
input node $v$ in $G$. The circuit evaluation phase of the protocol
for player $p$ starts after the completion of the \maj step for node
$v$ in \ic. After this step, for each input node $v$, players in $Q_v$
know the masked input $\mv_v$ and each has a share of the random
element $\mask_v$, although the actual input and mask are unknown to
any single player.
\begin{figure}[t]
  \begin{center}
  \includegraphics[scale=0.26]{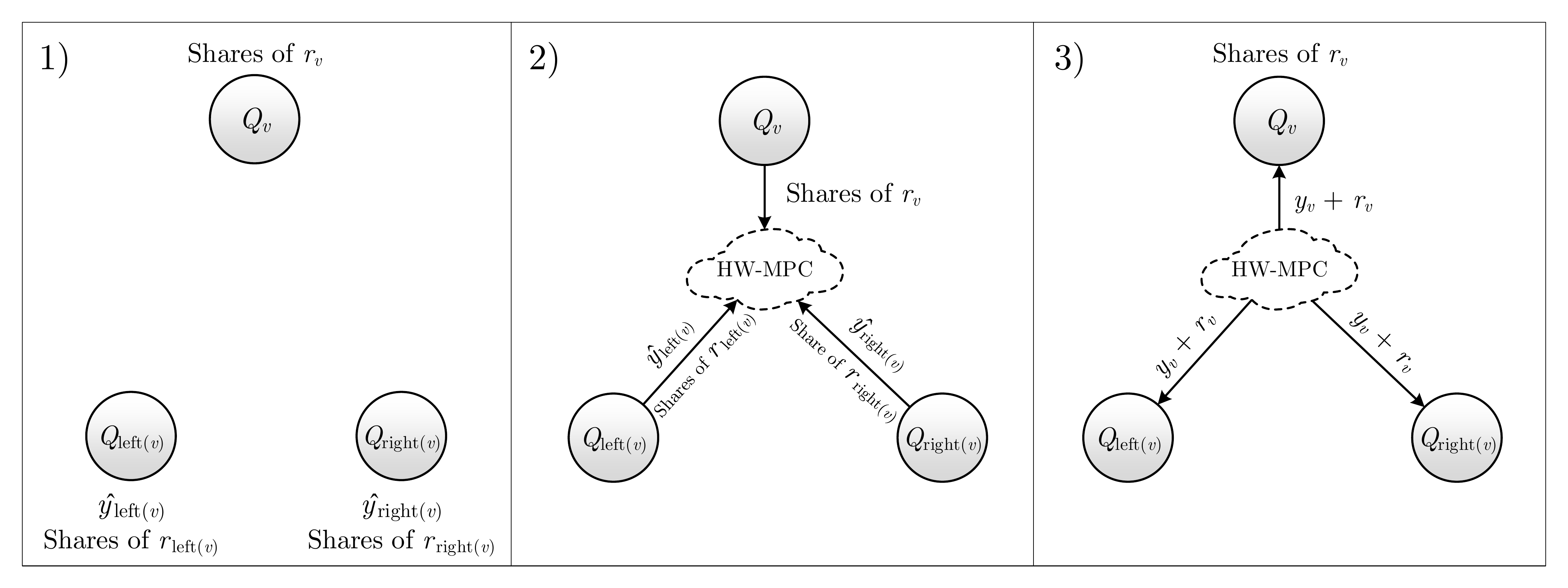}
  \end{center}
  \caption{Example Computation of a Gate associate with node $v$.}
  \label{f:quorums}
  \end{figure} 
The first step is to generate shares of uniformly random field
elements for all gate nodes. If player $p$ is in a quorum at gate node
$v$, he generates shares of $\mask_v$, a uniformly random field
element, by participating in the \gm algorithm. These shares are
needed as inputs to the subsequent run of \hw.

\begin{algorithm}
\caption{\ce}\label{alg:ce}
Run by each player $p$ in a quorum associated with each node $v \in G$
\begin{enumerate}
\item For each input node $v$, after finishing \ic with $\mv_v$ and a share of $\mask_v$ as output, $p$ uses these as inputs to \gcomp on each parent node of $v$.
\item For each gate node $v$:
\begin{enumerate}
\item $p$ runs \gm on $v$ and gets a share of $\mask_v$ as output
\item $p$ runs \gcomp on $v$ with its share of $\mask_v$ as the input and gets $\mv_v$ as output
\item $p$ runs \gcomp on each parent node of $v$ with input $\mv_v$ and $p$'s share of $\mask_v$
\end{enumerate}
\item After finishing computation of the gate represented by the output node, the players at the output node run \outrec to reconstruct the output.
\end{enumerate}
\end{algorithm}

\begin{algorithm}
\caption{\gm}\label{alg:rand}
This protocol is run by each player $p$ in a quorum associated with each gate node $v \in G$ to generate $\mask_v$.
\begin{enumerate}
\item Choose uniformly at random an element $r_{p,v} \in \F$ (this must be done independently each time this algorithm is run and independently of all other randomness used to generate shares of inputs etc.)
\item Run \vssSh to create verifiable secret shares of $r_{p,v}$ for each player in the quorum associated with $v$ and deal these shares to all the players in the quorum associated with $v$ including itself.
\item Participate in the \vssSh verification protocol for each received share. If the verification fails, set the particular share value to zero. 
\item Add together all the shares (including the one dealt by yourself).  This sum will be player $p$'s share of the value $\mask_v$. 
\end{enumerate}
\end{algorithm}

Next, players form the gadget for each gate node $v$ with children
$\lef(v)$ and $\rit(v)$ to evaluate the gate associated with $v$
using~\gcomp as depicted in Figure~\ref{f:quorums}. The values
$\vl_{\lef(v)}$ and $\vl_{\rit(v)}$ are the inputs to the gate
associated with $v$, and $\vl_v$ is the output of $v$ as it would be
computed by a trusted party. First section of figure describes the
initial conditions of the gate quorum and two input quorums before
participating in \hw. Each player in gate quorum $Q_v$ has a share of
the random element $\mask_v$ (via \gm). Every player in the left input
quorum $Q_{\lef(v)}$ has the masked value $\mv_{\lef(v)} =
\vl_{\lef(v)} + \mask_{\lef(v)}$ and a share of $\mask_{\lef(v)}$
(resp. for the right input quorum). In the second section, all the
players of the three quorums run \hw, using their inputs, in order to
compute $\mv_v$, which is equal to $\vl_v + \mask_v$. Third section shows
the output of the gate evaluation after participating in \hw. Each
player in $Q_v$ now knows 1) the output of the gate plus
the value of $\mask_v$; and 2) shares of $\mask_v$. Thus, players in
$Q_v$ now have the input to perform the computation associated with
the parents of $v$ (if any). Note that both $\vl_v$ and $\mask_v$
themselves are unknown to any individual.

The gate evaluation is performed for all gate nodes from the bottom of
the $G$ to the top. The output of the quorum associated with the
output node in $G$ is the output of the entire algorithm. Thus, this
quorum will unmask the output via \outrec. The last step of the
algorithm is to send this output to all players. We do via
a complete binary tree of quorums, rooted at the output quorum.

\begin{algorithm} 
\caption{\gcomp}\label{alg:gate}
This protocol is run for each gate node $v$ with children $\lef(v)$ and $\rit(v)$, the 
participants are the players in $Q_v$, $Q_{\lef(v)}$ and $Q_{\rit(v)}$.

\begin{enumerate}
\item If you are a player in  $Q_{\lef(v)}$, (resp. $Q_{\rit(v)}$) use $(\mv_{\lef(v)}, \mbox{ share of } \mask_{\lef(v)})$  (resp. $(\mv_{\rit(v)}, 
\mbox{ share of } \mask_{\rit(v)})$) as your input to \hw. If you are a player in $Q_v$, use $\mbox{ share of } \mask_v$ as your input to \hw.
\item Participate in \hw.
\item $\mv_v \leftarrow $ value returned by \hw.
\end{enumerate}
\end{algorithm}

\begin{algorithm} 
\caption{\outrec}\label{alg:root}
This protocol is run by all players in $Q_{\outnodemath}$.
\begin{enumerate}
\item Reconstruct $\mask_{\outnodemath}$ from its shares using \vssRec. 
\item Set the circuit output message: $\outval \leftarrow \mv_{\outnodemath} - \mask_{\outnodemath}$.
\item Send $\outval$ to all players in the quorums numbered $2$ and $3$.
\end{enumerate}
\end{algorithm}


\section{Some Remarks} 
\label{s:dis}
As described in the introduction, the goal of MPC is to simulate a
trusted third party in computation of the circuit and then send back
the computation result to the players.  Let $S$ denote the set of
players from whom input is received by the (simulated) trusted
party. Recall that $|S| \ge n-t$.\footnote{We allow $|S| >
  n-t$ because the adversary is not limited to delivering one message
  at a time; two or more messages may be received simultaneously.}
Thus, for an arbitrary $S$ a description of $S$ requires $\Omega(n)$
bits, and cannot be sent back to the players using only a scalable
amount of communication. Therefore, we relax the standard requirement
that $S$ be sent back to the players.  Instead, we require that at the
end of the protocol each good player learns the output of $f$; whether
or not their own input was included in $S$; and the \emph{size} of
$S$.

Also note that although we have not explicitly included this in \ic,
it is very easy for the players to compute the size of the computation
set $S$. Once each input quorum $Q_i$ has performed the \maj step and
agreed on the bit $b_i = \mathbf{1}_{i \in S}$ they can simply use an 
addition circuit to add these bits together and then
disperse the result. This is an MPC, all
of whose inputs are held by good players, since each input bit $b_i$
is jointly held by the entire quorum $Q_i$ and all the quorums are
good. Thus the computation can afford to wait for all $n$ inputs and 
computes the correct sum.

In the protocol proposed in this paper, it may be the case that a
player $p$ participates more than one time in the quorums performing a
single instance of \hw. In such a case, we allow $p$ to play the role
of more than one different players in the MPC, one for each quorum to
which $p$ belongs. This ensures that the fraction of bad players in
any instance of \hw is always less than $1/4$. \hw maintains privacy
guarantees even in the face of gossiping coalitions of constant
size. Thus, player $p$ will learn no information beyond the output and
its own inputs after running this protocol.

\section{Conclusion} \label{sec:conclusion}

We have described a Monte Carlo algorithm to perform asynchronous
secure multiparty computation in an scalable manner.  Our algorithms
are scalable in the sense that they require each player to send
$\tilde{O}(\frac{m}{n} + \sqrt n)$ messages and perform
$\tilde{O}(\frac{m}{n} + \sqrt n)$ computations.  They tolerate a
static adversary that controls up to a $\fbad-\epsilon$ fraction of
the players, for $\epsilon$ any positive constant.

Many problems remain open including the following. Can we prove lower
bounds for the communication and computation costs for Monte Carlo
MPC?  Can we implement and adapt our algorithm to make it practical
for a MPC problem such as the beet auction problem described
in~\cite{bogetoft2009secure}.  Finally, can we prove upper and lower
bounds for resource costs to solve MPC in the case where the adversary
is \emph{dynamic}, able to take over players at any point during the
algorithm?

\section{Acknowledgments}

This work was supported by the NSF under grants CCR-0313160 and CAREER Award 
0644058. We are grateful to Tom Hayes, Mahdi Zamani and the anonymous 
reviewers for useful suggestions and comments.


\bibliographystyle{abbrv}
\bibliography{security}

\end{document}